\documentclass[aps,superscriptaddress,twocolumn,preprintnumbers,showpacs,amsmath]{revtex4-2}

\usepackage{graphicx}
\usepackage{graphics}
\usepackage{dcolumn}
\usepackage{bm}
\usepackage{epstopdf}
\usepackage{mathrsfs}
\usepackage{amssymb}
\usepackage{amsmath}
\usepackage{esint}
\usepackage[pdftex]{hyperref}
\usepackage{xspace}
\usepackage{slashed}
\usepackage{nicefrac}
\usepackage{xcolor}
\usepackage[normalem]{ulem}
\usepackage{physics}

\def\p{{\boldsymbol p}}

\def\x{{\boldsymbol x}}

\def\d{\text{d}}
\def\rmd{\text{d}}
\def\rme{\text{e}}

\newcommand{\smallG}{{\scriptscriptstyle{G}}}
\newcommand{\smallS}{{\scriptstyle{S}}}
\newcommand{\smallGZ}{{\scriptscriptstyle{GZ}}}

\newcommand{\beq}{\begin{eqnarray}}
\newcommand{\eeq}{\end{eqnarray}}
\newcommand{\be}{\begin{eqnarray*}}
\newcommand{\ee}{\end{eqnarray*}}
\newcommand{\bqa}{\begin{eqnarray}}
\newcommand{\eqa}{\end{eqnarray}}

\newcommand{\nn}{\nonumber\\ }

\begin{document}

\title{Spectral sum rules and phase transition in strongly coupled QCD}

\author{Yi-Lun Du}
\email{yilun.du@iat.cn}
\affiliation{%
Shandong Institute of Advanced Technology, Jinan 250100, China}

\author{Nan Su}
\email{banban.su@gmail.com}
\noaffiliation

\author{Konrad Tywoniuk}
\email{konrad.tywoniuk@uib.no}
\affiliation{%
Department of Physics and Technology, University of Bergen, 5020 Bergen, Norway}

\date{\today}

\begin{abstract}

By incorporating contributions from both the (chromo)electric scale $gT$ and (chromo)magnetic scale $g^2T$, we establish spectral sum rules of quarks for strongly coupled QCD that respect Fermi-Dirac statistics as required by quantum mechanics. In sharp contrast to QED and weakly coupled QCD whose spectral functions consist of discontinuous zero-dimensional (poles) and one-dimensional (branch cuts) non-analytic contributions from real energy $p_0 \in \mathbb{R}$, the derived spectral function for strongly coupled quarks features continuous but non-analytic contributions from complex energy $p_0 \in \mathbb{C}$ that are two-dimensional in nature. In light of the novel sum rules, we uncover an intrinsic QCD transition between a three-mode phase at small coupling and a one-mode phase at large coupling. The transition is induced by the magnetic scale that generates a massless hydro-like mode with the genuine non-Abelian feature of positivity violation and serving as the Goldstone mode of the Lorentz symmetry breaking. The thermal mass serves as an order parameter of the transition and vanishes at large coupling in line with phenomenological predictions from Dyson-Schwinger equations and gauge/gravity duality. This result provides novel insights into the mechanism of the QCD deconfinement transition.

\end{abstract}

\maketitle

\section{Introduction}
\label{sec:intro}

Understanding the deconfinement transition of quantum chromodynamics (QCD) and the associated dynamical behavior of strongly coupled QCD matter is one of the most significant open questions in high energy physics~\cite{Alkofer:2006fu,Shuryak:2014zxa,Ratti:2021ubw}. As relativistic heavy-ion collisions at RHIC and LHC have demonstrated over the last two decades, the QCD matter generated in the experiments behaves as a strongly coupled system or, in other words, a liquid characterized by a small viscosity over entropy density ratio \cite{Gale:2013da,Heinz:2013th,Busza:2018rrf}. This challenges the theoretical description of the system based on conventional thermal QCD methods, suggesting a potential inconsistency of such approaches~\cite{Shuryak:2014zxa}. A first-principle description of dynamical QCD processes in the strongly coupled regime remains a daunting challenge.

The spectral function $\rho(\omega,p)$ is a fundamental quantity in the determination of physical degrees of freedom and transport properties, see e.g.~\cite{Casalderrey-Solana:2011dxg}. In contrast to the well-understood system of thermal quantum electrodynamics (QED), the spectral function of thermal QCD remains a key challenge~\cite{Fischer:2018sdj,Dupuis:2020fhh}. For a quantum many-body system, such as QED and QCD, canonical anti-commutation relations of the fermions $\psi(x)$ directly imply a sum rule for the corresponding fermionic spectral function $\rho(\omega,p)$. At finite temperature, it reads
\begin{multline}
    \label{eq:sumrule-qm}
    \int_{-\infty}^\infty \frac{{\rm d} \omega}{2\pi} \rho(\omega,p) = \frac{1}{4} \int{\rm d}^4 x \,\delta(t)\, \rme^{-i \p \cdot \x} \\
    \times \langle \psi(x)\psi^\dagger(0) + \psi^\dagger(0)\psi(x) \rangle_\beta= 1 \,,
\end{multline}
where $\langle \ldots \rangle_\beta$ represents a thermal average. This spectral sum rule is derived at operator level \cite{Bellac:2011kqa} and must be respected at one-loop level which resums non-perturbatively thermal or statistical effects of the system. The one-loop sum rule serves as a crucial consistency check of the setup, ensuring that fermions in the system obey the canonical anti-commutation relations. Only higher-order quantum effects involving renormalization may break the spectral sum rule~\cite{Alkofer:2000wg}. It has been demonstrated that the spectral sum rule for electrons is precisely fulfilled at one-loop level, which guarantees a systematic description of weakly coupled electron-positron plasma by thermal QED~\cite{Bellac:2011kqa}. Due to the lack of knowledge of the quark spectral function, it is still unclear how to establish the one-loop spectral sum rule for thermal QCD.

The collective behavior of quarks and gluons gives rise to two distinct thermal scales, namely the (chromo)electric scale $gT$ and the (chromo)magnetic scale $g^2T$, where $g$ is the coupling strength and $T$ the temperature. The absence of an infrared (IR) cutoff in the magnetic sector causes the breakdown of a perturbative expansion of finite-temperature Yang-Mills theory at the magnetic scale, often referred to as the Linde problem~\cite{Linde:1980ts,Gross:1980br}. The non-perturbative nature of the magnetic scale is intimately related to the confining properties of the dimensionally reduced Yang-Mills theory at high temperature~\cite{Buchmuller:1994qy,Alexanian:1995rp,Jackiw:1995nf,Jackiw:1997jga,Cornwall:1997dc}. As a result, conventional thermal QCD methods are incapable of incorporating the magnetic scale~\cite{Blaizot:2001nr,Kraemmer:2003gd,Andersen:2004fp,Arnold:2007pg,Su:2012iy}. It follows that the collective excitations of thermal QED and conventional thermal QCD are qualitatively the same, i.e. they are quasi-particles with thermal masses on the order of $gT$ that generate short-range correlations. Due to the missing of long-range correlations in the setup of conventional thermal QCD, it describes a weakly coupled plasma, just as that of thermal QED, in contrast to the strongly coupled liquid behavior observed in heavy-ion experiments \cite{Busza:2018rrf}.

The Gribov-Zwanziger (GZ) quantization of Yang-Mills theories, a formalism to study color confinement, is a promising method to tackle these issues~\cite{Gribov:1977wm,Zwanziger:1989mf}. It regulates the IR behavior of Yang-Mills theories by fixing the residual gauge transformations that remain after applying the Faddeev-Popov gauge-fixing procedure. This process results in a renormalizable action, namely the GZ action, which provides a systematic framework for analytic calculations. The gluon propagator in general covariant gauge of the GZ action reads
\begin{equation}
    \label{eq:gluon-propagator}
    D^{\mu\nu}(P) = \left[ \delta^{\mu\nu} - (1-\xi)\frac{P^\mu P^\nu}{P^2} \right] \frac{P^2}{P^4 + \gamma_\smallG^4} \;,
\end{equation}
where $\xi$ is the gauge parameter. Here $\gamma_\smallG$ is the Gribov parameter which is solved self-consistently from a gap equation that is defined to infinite loop orders, see Refs.~\cite{Dokshitzer:2004ie,Vandersickel:2012tz} for reviews. The GZ gluon propagator is IR suppressed, manifesting confinement effects, see Ref.~\cite{Maas:2011se} for a review, and is thus a major improvement over the conventional one from the Faddeev-Popov quantization that is adopted in conventional methods. 

More recently, this formalism has been generalized to finite-temperature \cite{Zwanziger:2004np,Zwanziger:2006sc,Fukushima:2013xsa,Cooper:2015bia}. In this case, the one-loop gap equation can be solved analytically at asymptotically high temperatures and gives~\cite{Zwanziger:2006sc,Fukushima:2013xsa}
\begin{equation}
    \label{eq:gamma}
    \gamma_\smallG = \frac{D-1}{D} \frac{N_c}{4\sqrt{2}\pi} g^2T \,,
\end{equation}
where $D$ is the space-time dimensions and $N_c$ is the number of colors. It is noted that Eq.~\eqref{eq:gamma} provides a fundamental IR cutoff at the non-perturbative magnetic scale for the GZ action. The resulting equations of state are significantly improved over the ones from conventional resummed perturbation theory~\cite{Zwanziger:2004np,Fukushima:2013xsa,Canfora:2015yia}.

Collective excitations are effective degrees of freedom of a many-body system that are crucial in determining its physical properties. A key finding from the GZ quantization is that the magnetic scale generates a massless collective excitation behaving like the sound wave in hydrodynamics and, surprisingly, this long-range mode induces positivity violation indicating the indispensability of confinement effects in the establishment of a strongly coupled description of thermal QCD~\cite{Su:2014rma}. This result combines the expectation of massless mode from quark-gluon plasma phenomenology~\cite{Hidaka:2011rz,Satow:2013oya,Blaizot:2014hka,Policastro:2002tn,Kovtun:2005ev,Kitazawa:2005mp,Kitazawa:2006zi,Kitazawa:2007ep,Qin:2010pc,Gao:2014rqa,Wang:2018osm,Tripolt:2020irx} and positivity violation from color confinement studies~\cite{Cucchieri:2004mf,Sternbeck:2006cg,Bowman:2007du,Silva:2013rrr,Alkofer:2003jj,Maas:2004se,Maas:2005hs,Fischer:2008uz,Fister:2011uw,Strauss:2012dg,Karsch:2009tp,Mueller:2010ah,Qin:2013ufa} into one systematic scheme that is intrinsically strongly coupled. 

In this paper, we present a first study of the spectral function of thermal quarks incorporating effects from both the electric and magnetic scales via the GZ quantization. It is well-known that spectral function reflects the non-analytic structure of the system. The non-analytic structure of the quark propagator shows striking features in the GZ quantization: unlike QED whose non-analytic behavior is located exclusively on the real energy axis, there are non-analytic regions in the complex plane besides the real energy axis in the QCD case. To capture this behavior, we define a novel spectral density sensitive to the local non-analytic features in the complex plane. The resulting thermal quark spectral function represents a strongly coupled system characterized with confinement effects via positivity violation. Furthermore, we establish the spectral sum rule of thermal quarks at one-loop order to demonstrate the consistency of the GZ quantization, in which the non-analytic regions in the complex plane play an indispensable role. Finally, we derive an intrinsic non-Abelian phase transition of the system induced by magnetic scale in light of the spectral sum rules: with increasing coupling strength the system transforms from a three-mode phase (two massive quasi-particle modes and one massless, hydrodynamic-like mode) to a one-mode phase consisting of quarks in the vacuum. Hence, in the one-mode phase Lorentz symmetry is restored, as required by self-consistency.

The paper is organized as the following: in Sec.~\ref{sec:sum-rules} we analyze the structure of the strongly coupled thermal quark propagator in the complex energy plane and derive the corresponding spectral sum rules. In Sec.~\ref{sec:transition-p0}, we uncover an intrinsic QCD transition at vanishing spatial momentum in light of the spectral sum rule that features Lorentz symmetry breaking. In Sec.~\ref{sec:transition-pfinite}, we move on to analyze the transition at finite spatial momentum. We finally conclude in Sec.~\ref{sec:discussion}. The novelty of the results arises exclusively due to the nature of the complex conjugate poles in Eq.~\eqref{eq:gluon-propagator}, which is a shared feature of all Gribov-like approaches. We thus use the GZ action as a simple demonstration, without the loss of generality, though the refined GZ action is in better agreement with lattice data~\cite{Dudal:2008sp}. 

\section{Spectral sum rules for strongly coupled QCD}
\label{sec:sum-rules}

The basis for deriving the spectral properties of quark in a thermal medium is its propagator. With $P=(p_0,\p)$ the four-momentum and $\gamma_\mu=(\gamma_0,{\boldsymbol\gamma})$ the gamma matrices, we define the Minkowskian quark propagator as
\begin{equation}
    \label{eq:ResumProp}
    i S^{-1}(P) = \slashed P - \Sigma(P) \equiv A_0 \gamma_0 - A_\smallS \hat \p\cdot {\boldsymbol\gamma} \,,
\end{equation}
where $\Sigma(P)$ is the self-energy. The propagator can be decomposed into positive and negative helicity-to-chirality contributions~\cite{Bellac:2011kqa}. For the one-loop quark propagator in the GZ quantization, and within the hard-thermal-loop (HTL) systematics \cite{Klimov:1982bv,Weldon:1983jn,Braaten:1989mz}, see also \cite{Su:2014rma}, we obtain
\begin{align}
\label{eq:A0}
    A_0(p_0,p) &= p_0 - \frac{g^2C_F}{2(2\pi)^2}\sum_\pm\int_0^\infty\d k\, k\,\tilde n_\pm(k,\gamma_\smallG) \nn 
    &\;\;\;\;\;\times \left[Q_0(\tilde\omega^\pm_1,p) + Q_0(\tilde\omega^\pm_2,p) \right] \,,\\
\label{eq:AS}
    A_\smallS(p_0,p) &= p + \frac{g^2C_F}{2(2\pi)^2} \sum_\pm\int_0^\infty \d k \,k\, \tilde n_\pm(k,\gamma_\smallG) \nn 
    &\;\;\;\;\;\times \left[Q_1(\tilde\omega^\pm_1,p) + Q_1(\tilde\omega^\pm_2,p) \right] \,,
\end{align}
where $C_F \equiv (N_c^2-1)/(2N_c)$ is the quadratic Casimir operator in the fundamental representation, $\tilde n_\pm(k,\gamma_\smallG)\equiv n_B(\sqrt{k^2 \pm i \gamma_\smallG^2}) + n_F(k)$ with $n_B$ and $n_F$ the Bose-Einstein and Fermi-Dirac distributions respectively, $\tilde\omega^\pm_1 \equiv E_\pm(p_0 + k - E_\pm)/k$ and $\tilde\omega^\pm_2 \equiv E_\pm(p_0 - k + E_\pm)/k$ with $E_\pm=\sqrt{k^2 \pm i\gamma_\smallG^2}$, and the Legendre functions $Q_0(\omega,p) \equiv \frac{1}{2 p}\ln \frac{\omega + p}{\omega - p}$ and $Q_1(\omega,p)\equiv (1-\omega Q_0(\omega, p))/p$~\cite{Su:2014rma}. We can easily reconstruct the propagator as
\begin{equation}
\label{eq:ResumProp}
    -i S(P) = \frac{\gamma_0 - \hat \p\cdot {\boldsymbol\gamma}}{2} \Delta_+(P) + \frac{\gamma_0 + \hat \p\cdot {\boldsymbol\gamma}}{2} \Delta_-(P) \,,
\end{equation}
where $\Delta_\pm (P) = (A_0(p_0,p) \mp A_\smallS(p_0,p))^{-1}$. Here we have made explicit that the propagator only depends on the magnitude of the three-momentum $p \equiv |\p|$. The effective propagator $\Delta_+(P)$ can now be written as
\begin{equation}
\label{eq:qcd-propagator}
    \Delta_+(P) = \frac{1}{p_0 - p - \tilde \Sigma(P)} \,.
\end{equation}
Here we have defined the self-energy as
\begin{multline}
\label{eq:sigma-tilde}
    \tilde \Sigma(p_0,p) = \frac{g^2C_F}{2(2\pi)^2}\sum_\pm \int_0^\infty \rmd k\, k \,\tilde n_\pm(k,\gamma_\smallG) \\ \times \big[{\cal Q}(\tilde \omega_1^\pm,p) + {\cal Q}(\tilde \omega_2^\pm,p) \big] \,,
\end{multline}
where we introduced the short-hand ${\cal Q}(p_0,p) = Q_0(p_0,p) + Q_1(p_0,p)$. Completely analogous definitions can be found for $\Delta_-(P)$. For QED, with $\gamma_\smallG=0$, we simply get $\tilde \Sigma (p_0,p) = \frac{g^2 C_F}{2\pi^2} \int_0^\infty \rmd k \, k \tilde n(k) {\cal Q}(p_0,p)$.

To investigate the analytic structure of the propagator and, in particular, to study the analytic continuation from Euclidean to Minkowski metric, we shall consider $p_0 \in \mathbb{C}$ ($p_0 = x+iy$) in the following. First, we have the well-known parity properties of the propagator,
\begin{align}
    \Re \Delta_+(x+i y,p) &= - \Re \Delta_-(-x+iy,p) \,,\\
    \Im \Delta_+(x+iy,p) &= \Im \Delta_-(-x+iy,p) \,.
\end{align}
Hence it is sufficient to study $\Delta_+ (P)$ and its analytic properties to map out the analytic structure of the full propagator in Eq.~\eqref{eq:ResumProp} \cite{Bellac:2011kqa}. Furthermore, complex conjugate symmetry of the propagator, i.e. $\Delta^\ast_\pm(p_0,p) = \Delta_\pm\left( p_0^\ast,p\right)$, implies that
\begin{align}
    \Re \Delta_\pm(x+iy,p) &= \Re \Delta_\pm(x-iy,p) \,,\\
    \Im \Delta_\pm(x+iy,p) &= -\Im \Delta_\pm(x-iy,p) \,.
\end{align}
The propagator is therefore Hermitian analytic~\cite{Weldon:1983jn}.

The analytic structure of the propagator $\Delta_+(p_0,p)$ in the complex $p_0$ plane changes drastically between $\gamma_\smallG =0$ and $\gamma_\smallG >0$. We sketch the analytic structure in the two cases in Fig.~\ref{fig:complex-plane}. Let us also comment briefly on the used nomenclature. When $\gamma_\smallG=0$, we refer to it as the QED or weakly coupled QCD case. The system only contains contributions from the electric scale (and the main difference between QED and QCD is the corresponding color factor $C_F$ appearing for the latter). The $\gamma_\smallG >0$ case is referred to as strongly coupled QCD, or simply QCD, throughout, where both electric and magnetic scales are at play.

\subsection{Sum rules for QED}
\label{sec:sum-rules-qed}

In QED (or weakly coupled QCD) the propagator is known analytically,
\begin{equation}
\label{eq:qed-propagator-analytic}
    \Delta_\pm (P) = \frac1{p_0 \mp p - \frac{m_q^2}{2p}\left[ \left(1\mp \frac{p_0}{p} \right)\ln\frac{p_0 + p}{p_0 - p} \pm 2\right]} \,,
\end{equation}
where $m_q^2 = \frac{2g^2 C_F}{(2\pi)^2}  \int_0^\infty \rmd k\, k [n_B(k)+n_F(k)] =  g^2 T^2C_F/8$ is the thermal quark mass. Technically, we should also remove the additional color degrees of freedom by replacing $C_F g^2 \to e^2$ to obtain the correct QED results \cite{Bellac:2011kqa}. 

The function is holomorphic in the complex plane except on the real axis, where $p_0 \equiv \omega$ with $\omega \in \mathbb{R}$. In addition to two poles at $\omega = \omega_\pm(p)$ there is a branch cut $\omega\in(-p, p)$ in the self-energy, displaying the weakly coupled nature of the underlying system~\cite{Bellac:2011kqa}. The poles are interpreted in linear response theory as quasi-particles of the system, a dressed quark and an anti-quark ``hole'' excitation in the medium. The cut, on the other hand, is a feature of Landau damping of these quasi-particles.

\begin{figure*}
    \centering
    \includegraphics[width=0.35\textwidth]{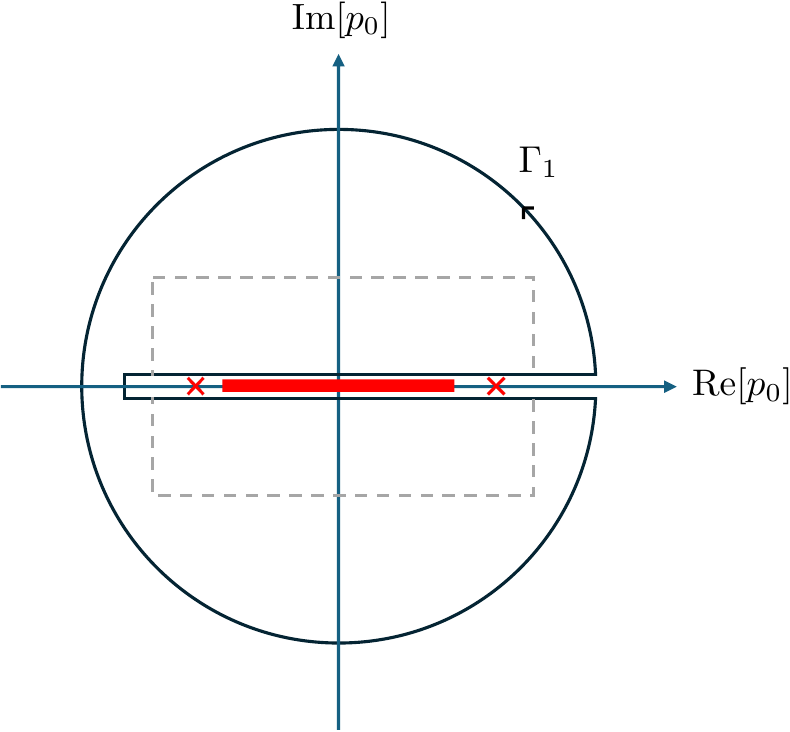}
    \hspace{1em}
    \includegraphics[width=0.35\textwidth]{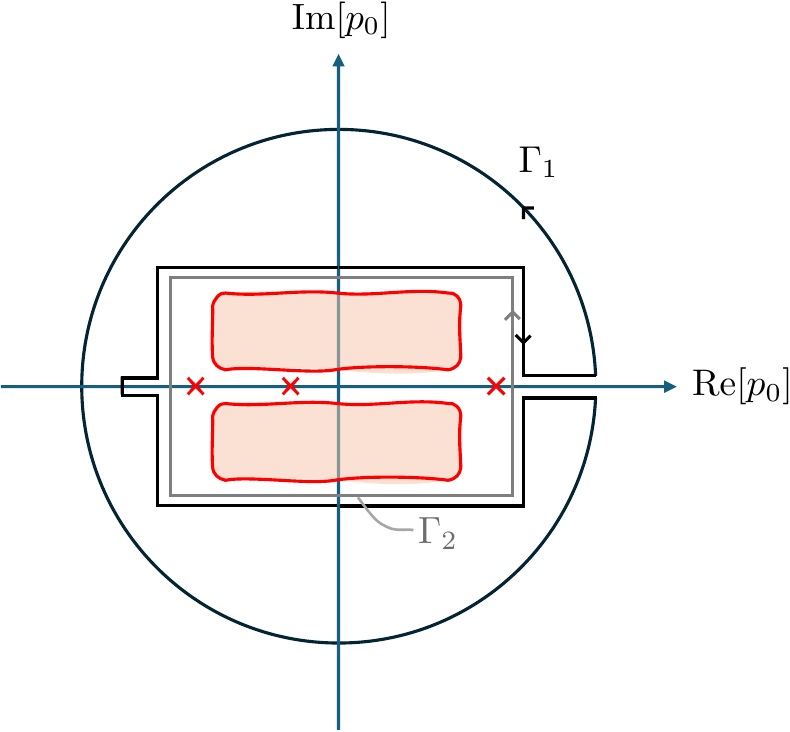}
    \caption{The analytic structure of the propagator $\Delta_+(p_0,p)$ in the complex-$p_0$ plane is shown for QED (left panel) and QCD (right panel). The dashed gray line in the left panel highlights the contrast between the two cases. Red crosses represent poles, the continuous red line indicates a branch cut, and the red regions denote smooth yet non-analytic features.}
    \label{fig:complex-plane}
\end{figure*}

To better illustrate the differences between QED and QCD, let us briefly recall the derivation of sum rules in the former theory. For deriving sum rules, we need the structure of the propagator in large-$|p_0|$ limit. Given the simple analytic structure of the QED propagator we can avoid any non-holomorphic feature in the complex plane by a clever choice of the contour, and consequently write Cauchy's integral formula as
\begin{equation}
\label{eq:cauchy-integral}
    \Delta_\pm(p_0,p) = \ointctrclockwise_{\Gamma_1} \frac{\rmd z}{2\pi i} \frac{\Delta_\pm(z,p)}{z-p_0} \,,
\end{equation}
where the contour $\Gamma_1$ is depicted in Fig.~\ref{fig:complex-plane} (left panel). The contour along the real axis is deformed into the complex $p_0$ plane with a deviation $i \eta$, to obtain
\begin{align}
    \Delta_\pm(p_0,p) &= \int_{-\infty}^{\infty} \frac{\rmd z}{2\pi i} \frac{\Delta_\pm(z+i \eta,p) - \Delta_\pm(z-i \eta,p)}{z-p_0} \nonumber \\ 
    &+ \ointctrclockwise_{\Gamma'} \frac{\rmd z}{2\pi i} \frac{\Delta_\pm(z,p)}{z} \,,
\end{align}
where $\Gamma'$ is the circular contour with radius tending to infinity. Since $\Delta_\pm(z,p) \sim 1/z$ for $z \to \infty$, we can drop the second term entirely. Then, using the Hermitian property of the propagator and the definition of the spectral function $\rho(p_0,p) = -2 \Im \Delta_\pm(p_0,p)$ \footnote{Note the sign indicating a fermionic spectral function}, we obtain
\begin{equation}
    \Delta(p_0,p) = -\int_{-\infty}^\infty \frac{\rmd z}{2\pi} \frac{\rho(z,p)}{z-p_0} \,.
\end{equation}
Here we have taken the $\eta \to 0$ limit. Expanding both sides of this equation in inverse powers of $p_0$ at the large $p_0$ limit, we obtain the sum rules \cite{Bellac:2011kqa}
\begin{subequations}
\label{eq:sumrules_n_QED}
\begin{align}
    \int_{-\infty}^\infty \frac{\rmd z}{2\pi} \,\rho_\pm(z,p) &= 1\,,\label{eq:qed-1} \\
    \int_{-\infty}^\infty \frac{\rmd z}{2\pi} \,z \rho_\pm(z,p) &= \pm p \,,\\
    \int_{-\infty}^\infty \frac{\rmd z}{2\pi} \,z^2 \rho_\pm(z,p) &= p^2 + m_q^2 \,.
\end{align}
\end{subequations}
Since the non-analytic features of the propagator have been fully captured on the real axis, the spectral function contains all physical information. Note that the sum rule in Eq.~(\ref{eq:qed-1}) confirms that electrons obey the anti-commutation relation as required by quantum mechanics in Eq.~(\ref{eq:sumrule-qm}),  which serves as a nontrivial self-consistency check for thermal QED~\cite{LeBellac}.

\subsection{Sum rules for QCD}
\label{sec:sum-rules-qcd}

The analytic structure of $\Delta_+(\gamma_\smallG)$ in the strongly coupled QCD case is considerably more involved than the QED or weakly coupled  QCD case, and it thus should be treated with more care, see Fig.~\ref{fig:complex-plane} (right panel). In contrast to the QED case, the imaginary part of the propagator is exactly zero along the real axis \cite{Su:2014rma}. Furthermore, we encounter regions in the complex plane that are smooth but non-analytic in addition to discontinuous cuts there, as will be presented in the following sections.

Next, we introduce sum rules for strongly coupled QCD, which can be used to study the properties of the propagator at general couplings $g$. Same as for the QED case, our goal is to extract information in the large $p_0$ limit. Since the non-analytic features are located close to the origin of the complex-$p_0$ plane, we can use the Cauchy formula Eq.~\eqref{eq:cauchy-integral} for the $\gamma_\smallG >0$ case as well. However, in this case, the contour $\Gamma_1$ will have to avoid a larger region near the origin.

Similar as in QED, we expand the factor in the denominator on the right-hand-side of Eq.~(\ref{eq:cauchy-integral}) at the large $p_0$ limit, arriving at
\begin{equation}
    \Delta_\pm(p_0,p) = - \frac{1}{p_0} \sum_{n=0}^\infty \ointctrclockwise_{\Gamma_2} \frac{\d z}{2\pi i}\, \left(\frac{z}{p_0} \right)^n \Delta_\pm(z,p) \,,
\end{equation}
where the contribution of a circular contour $\Gamma'$ with radius tending to infinity vanishes as in QED. By expanding the resummed propagator $\Delta_\pm(p_0, p)$ in Eq.~\eqref{eq:qcd-propagator} in negative powers of $p_0$ on the left-hand side and comparing both sides, we obtain for $n=0, 1$ and $2$, respectively,
\begin{subequations}
\label{eq:sumrules_n}
\begin{align}
\label{eq:sumrule_n=0}
    \ointclockwise_{\Gamma_2} \frac{\d z}{2\pi i}  \,\Delta_\pm(z,p) &= 1 \,,\\
\label{eq:sumrule_n=1}
    \ointclockwise_{\Gamma_2} \frac{\d z}{2\pi i} \, z \Delta_\pm(z,p) &= \pm p \;,\\
\label{eq:sumrule_n=2}
    \ointclockwise_{\Gamma_2} \frac{\d z}{2\pi i} \, z^2 \Delta_\pm(z,p) &= p^2+ \tilde m_q^2(\gamma_\smallG) \;.
\end{align}
\end{subequations}
Note that the contour integral is oriented in the opposite direction now. The parameter in the $n=2$ sum rule is defined as $\tilde m_q^2(\gamma_\smallG) = \frac{g^2 C_F}{4\pi} \sum_\pm \int_0^\infty \rmd k \,k^2 \tilde n_\pm(k,\gamma_\smallG)/E_\pm$. In the $\gamma_\smallG=0$ limit, this reduces to the conventional quark screening mass $\tilde m_q(0) =m_q$. For strongly coupled QCD, this definition deviates from the real mass scale, which tends to vanish at large couplings, see Sec.~\ref{sec:mass}.

We see that for $n\leq 2$, the terms from expanding $\Delta_\pm(p_0, p,\gamma_\smallG)$ look identical to the QED case, except that $\tilde m_q(\gamma_\smallG)$ captures the contributions from the magnetic scale $g^2T$ at $n=2$. However, for $n\geq 3$, the terms from expanding $\Delta_\pm(p_0, p,\gamma_\smallG)$ will no longer be polynomials of $p$ and $\tilde m_q(\gamma_\smallG)$ as the QED case, which is attributed to the complex behavior of the magnetic scale at higher orders.

The sum rules are of crucial importance to control the behavior and self-consistency of the system when varying the coupling strength $g$, guaranteeing that all necessary degrees of freedom are accounted for. For QED, the sum rules in Eqs.~\eqref{eq:sumrules_n_QED} are trivially respected at any coupling. For QCD, however, positivity violating contributions have to be treated with special care and the sum rules are a crucial consistency check.

The spectral sum rules also provide a way to constrain the possible non-analytic features in the complex energy plane. For this purpose, we define three \emph{partial} sum rules as
\begin{subequations}
\label{eq:partial-sum-rules}
\begin{align}
    \label{eq:partial-sum-rule-n0}
    \mathcal{S}^{(0)}(R,p) &= \ointctrclockwise_{\Gamma_R} \frac{\rmd z}{2\pi i} \, \Delta_+(z,p) \,,\\
    \mathcal{S}^{(1)}(R,p) &= \left\vert\frac1{p} \ointctrclockwise_{\Gamma_R} \frac{\rmd z}{2\pi i} \, z \Delta_+(z,p) \right\vert \,,\\
    \mathcal{S}^{(2)}(R,p) &= \frac{1}{p^2+\tilde m^2_q(\gamma_\smallG)} \ointctrclockwise_{\Gamma_R} \frac{\rmd z}{2\pi i} \, z^2 \Delta_+(z,p) \,,
\end{align}
\end{subequations}
where the contour $\Gamma_R$ is a circle with finite radius $R$ centered at the origin. With these definitions, when all non-analytic structures of $\Delta_+(p_0,p)$ are captured within the contour $\Gamma_R$, i.e., when $\Gamma_R$ acts as $\Gamma_2$, see Fig.~\ref{fig:complex-plane}, the sum-rules in Eqs.~\eqref{eq:sumrules_n} are exactly fulfilled. In other words, we have that $\mathcal{S}^{(i)}(\Gamma_R= \Gamma_2 ,p) =1$.

\subsection{Spectral density}
\label{sec:spectral-density}

As shown above, all of non-analytic structures in the complex $p_0$ plane can be captured by the contour $\Gamma_2$ in Fig.~\ref{fig:complex-plane} (right panel). Given the additivity of contour integrals over neighboring areas, in this subsection we aim to extract the local non-analytic structures. Consider a small contour $\gamma$ centered around some point $p_0$ in the complex plane and oriented counterclockwise. We use Green's theorem (the simplest version of Stoke's theorem) to write
\begin{equation}
\label{eq:contour-spectral}
    \ointctrclockwise_\gamma \frac{\d z}{2\pi i}  \,\Delta_\pm(z,p) = \int_S \frac{\rmd A}{2\pi} \,d_\pm(z,p) \,,
\end{equation}
where $S$ denotes the area covered by the contour $\gamma$, i.e. $\gamma = \partial S$, and $\rmd A = \rmd x \wedge \rmd y$ is the area element. This defines the \textit{spectral density} $d_\pm(p_0,p)$, as
\begin{equation}
\label{eq:spectral-density}
    d_\pm(z,p) = \frac{\partial u}{\partial x} - \frac{\partial v}{\partial y} + i \left(\frac{\partial u}{\partial y} + \frac{\partial v}{\partial x} \right) \,,
\end{equation}
where $u = \text{Re}\Delta_\pm$, $v=\text{Im}\Delta_\pm$ while $z = x+iy$. This can be rewritten as $d_\pm = 2\partial \Delta_\pm /\partial \bar z$, where the operator $\partial/\partial \bar z = \frac12 (\frac{\partial}{\partial x} + i \frac{\partial}{\partial y})$ is called the Wirtinger operator \cite{lars1990analysis}. The two derivative terms contained in the spectral density are the so-called Cauchy-Riemann conditions. Hence, $d_\pm(p_0,p) = 0$ implies an analytic (holomorphic) function in $p_0$, while the opposite immediately signals presence of non-analytic behavior.

Let us now choose the contour $\gamma$ to be a rectangle centered at $z$ with corners located at $x\pm\zeta$ and $y\pm\eta$. Now the contour integral can be rewritten as
\begin{multline}
\label{eq:sumrule_finite_spectral_n=0}
    \ointctrclockwise_\gamma \frac{\d z}{2\pi i}  \,\Delta_\pm(z,p) = \int_{x-\zeta}^{x+\zeta} \frac{{\d} x'}{2\pi}\, \rho_\eta (x'+iy,p) \\ + \int_{y-\eta}^{y+\eta}\frac{{\d} y'}{2\pi} \,\tilde \rho_\zeta(x+iy',p) \,.
\end{multline}
Here we have defined
\begin{equation}
\label{eq:spectral-function-eta}
    \rho_\eta(z,p) = i\big[\Delta_\pm(z +i\eta,p) - \Delta_\pm(z - i\eta,p) \big]\,,
\end{equation}
and
\begin{equation}
\label{eq:spectral-function-zeta}
    \tilde \rho_\zeta(z,p) = \Delta_\pm(z+ \zeta,p) - \Delta_\pm(z -\zeta ,p) \,.
\end{equation}
These definitions will be handy in two contexts. When considering small and local contours, we can derive a simple relation between the contour integral and the spectral density. For $\eta \to 0$ and $\zeta \to 0$, we can use the mid-point rule to approximate the right-hand side of Eq.~\eqref{eq:sumrule_finite_spectral_n=0}, as
\begin{equation}
    \ointctrclockwise_\Gamma \frac{\d z}{2\pi i}  \,\Delta_\pm(z,p) = \frac{\zeta}{\pi} \rho_\eta(z,p) + \frac{\eta}{\pi} \tilde \rho_\zeta(z,p) \,.
\end{equation}
Similarly, using the midpoint-rule on the right-hand side of Eq.~\eqref{eq:contour-spectral}, we finally obtain
\begin{equation}
\label{eq:spectral-density-practical}
    d_\pm(z,p) = \lim_{\eta \to 0}\frac{\rho_\eta(z,p)}{2\eta} + \lim_{\zeta \to 0}  \frac{\tilde \rho_\zeta(z,p)}{2\zeta} \,.
\end{equation}
This formula, equivalent to Eq.~(\ref{eq:spectral-density}), provides a fast and precise way of computing numerically the local spectral density at any point in the complex plane.

We can also make contact with the conventionally defined spectral function by considering points on the real axis $z=\omega \in \mathbb{R}$. In this case, we have
\begin{equation}
    \rho_\eta (\omega,p) = -2 \Im \Delta_\pm(\omega +i\eta,p) \,.
    \label{eq:rhoR-2d}
\end{equation}
Taking the limit $\eta \to 0$, we recover\textemdash as a consistency check\textemdash the conventional spectral function $\rho_\pm(\omega,p) = \lim_{\eta \to 0} \rho_\eta(\omega,p) = -2 \Im \Delta_\pm(\omega,p)$.

With the help of Green's theorem, we can also obtain 
\begin{equation}
    \ointctrclockwise_{\gamma}\frac{{\d}z}{2\pi i}\,z^n\Delta_\pm(z, p) = \int_{S}\frac{\rmd A}{2\pi}\,z^n d_\pm(z, p) \,,
    \label{eq:sumrule_density_recursive}
\end{equation}
which can be easily proven \footnote{In general, the contour integral of a product of two functions $f$ and $g$ becomes $\ointctrclockwise_{\partial S} \frac{\rmd z}{2\pi i} \, f(z) g(z) = \iint_S \frac{\rmd z\wedge \rmd \bar z}{\pi} \big[ \mathcal{D}(f) g + f \mathcal{D}(g) \big]$, where the operator $\mathcal{D}$ gives the two Cauchy-Riemann conditions, $\mathcal{D}(f)= \partial_x \Re f - \partial_y \Im f + i \big(\partial_y \Re f + \partial_x \Im f \big)$, with $z = x+iy$. For any analytic function $f$, $\mathcal{D}(f)=0$. According to our definitions above, $\mathcal{D}\big(\Delta_\pm(z,p)\big) = d_\pm(z,p)$. Furthermore, $\mathcal{D}(z^n \Delta_\pm) = z^n \mathcal{D}(\Delta_\pm)$, since $z^n$ is analytic in the whole complex plane.}. This suggests that the spectral density captures all relevant information about the non-analytic features of the propagator. Finally, comparing to the sum-rule in Eq.~\eqref{eq:sumrule_n=0}, i.e. replacing the contour $\gamma$ with $\Gamma$, we see that integrating the propagator over the contour corresponds to  integrating the spectral density over the corresponding area. 

\begin{figure*}
    \centering
    \includegraphics[width=\textwidth]{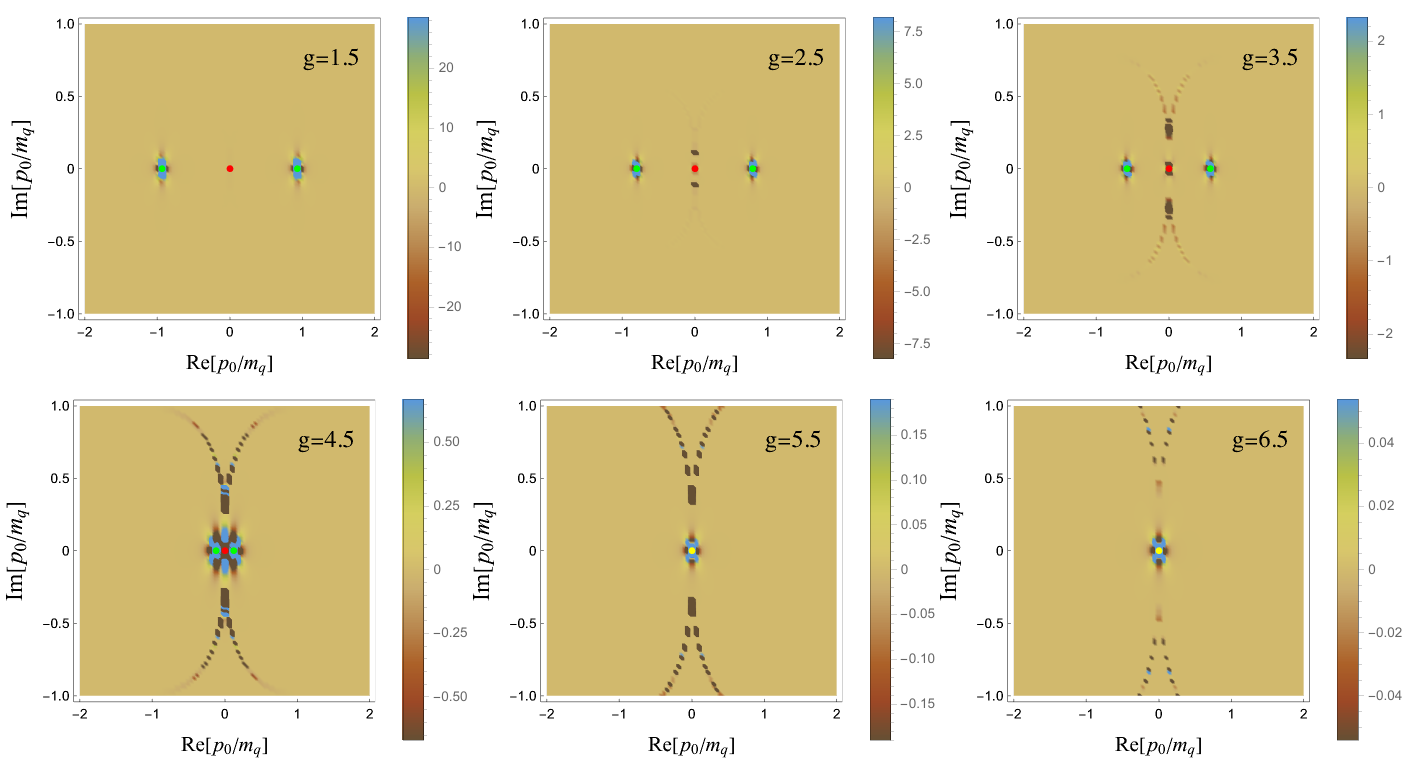}
    \caption{Plots of the real part of the spectral density in the static limit, $\Re d_+(p_0,0)$, for a range of coupling strengths. The locations of poles on the real axis have been marked with red (the GZ pole) and green (the plasmon and plasmino poles) dots at small couplings, while the unique pole at the origin is marked with white dot at large couplings.}
    \label{fig:transition-p0}
\end{figure*}

\section{QCD transition in the $p\to 0$ limit}
\label{sec:transition-p0}

In this section, we study the properties of the spectral function $d_\pm(p_0,p)$ at vanishing spatial momentum, i.e. $p \to 0$. 
We would like to emphasize that the sum rule in Eq.~(\ref{eq:sumrule_n=0}) is derived for arbitrary coupling, and is non-perturbative in nature due to its origin from quantum mechanics, despite the demonstration shown in Fig.~\ref{fig:complex-plane} (right panel) represents the case of small couplings. Therefore it enables us to explore the behavior of $d_\pm(p_0,p)$ at large couplings which is beyond the scope of conventional methods. We must make sure that Eq.~(\ref{eq:sumrule_n=0}) is satisfied at any stage, which is a highly non-trivial criterion. We demonstrate this explicitly in Sec.~\ref{sec:sum-rules-p0}.

The $p \to 0$ limit gives us access to the thermal masses of excitations of the strongly coupled QCD medium. In the absence of the GZ scale, i.e. $\gamma_\smallG = 0$, we get that
\begin{equation}
    \tilde \Sigma(\omega,0) = \frac{m^2_q}{\omega} \,,
\end{equation}
where the thermal mass $m_q$ is defined below Eq.~\eqref{eq:qed-propagator-analytic}. The pole of the propagator in Eq.~\eqref{eq:qcd-propagator} is therefore located at $\omega = m_q$. This corresponds to a static, screened excitation, equivalent to the QED fermionic thermal scale. 

\subsection{Transmutation of degrees of freedom}
\label{sec:dofs-p0}

While the analysis in QED is limited to real values of the energy, for QCD we extend the scope to study non-analytic features that are encoded in the spectral density at any $p_0 \in \mathbb{C}$ in the complex plane to map out the features of the propagator. We currently take the spatial momentum $p$ to zero, where the self-energy takes a simplified form,
\begin{equation}
    \tilde \Sigma(p_0,0) = \frac{g^2 C_F}{(2\pi)^2} \sum_\pm \int_0^\infty \rmd k\, \frac{\tilde n_\pm(k)\,k^2 \,p_0}{E_\pm \left[p_0^2 - (E_\pm-k)^2\right]} \,.
\end{equation}
This allows us to compute the spectral function $d_+(p_0,0)$ using Eq.~\eqref{eq:spectral-density-practical} together with Eq.~\eqref{eq:spectral-function-eta} and Eq.~\eqref{eq:spectral-function-zeta}.

We have plotted $\Re d_+(p_0,0)$ in the complex plane of $p_0/m_q$ (where $m_q$ is the QED mass as a scale) for several values of the coupling $g$ in Fig.~\ref{fig:transition-p0}. Note that imaginary part of spectral density $\Im d_+(p_0,p)$ makes no contribution to the sum rule for $n=0$. The range of couplings spans $1.5 < g < 6.5$. The temperature scales out in any relevant quantity through the temperature-dependent parameters $m_q$ and $\gamma_\smallG$. We have also chosen $\eta$ and $\zeta$ sufficiently small in Eq.~\eqref{eq:spectral-density-practical} to reach numerically robust results. 

Figure~\ref{fig:transition-p0} contains several remarkable features. Firstly, we see that the spectral density is non-zero in the complex plane, not only on the real axis. Secondly, we observe a transition between a three-pole structure at small coupling to a unique pole at large coupling. 

\begin{figure}
    \centering
    \includegraphics[width=\columnwidth]{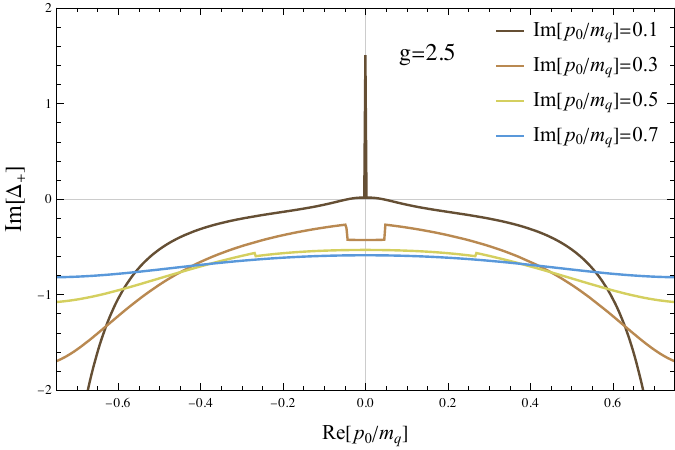}
    \includegraphics[width=\columnwidth]{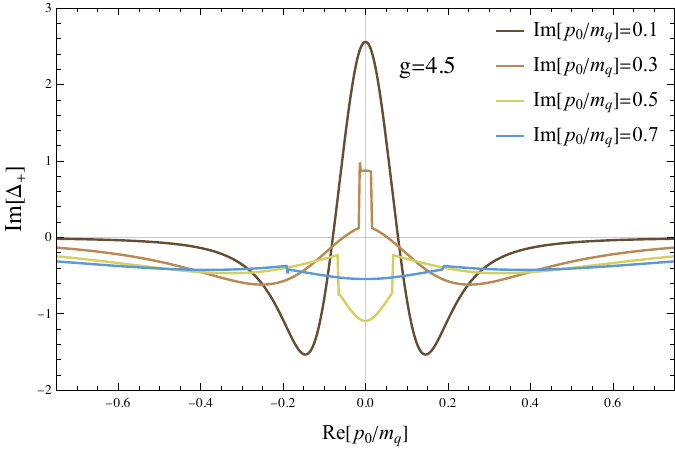}
    \caption{Discontinuities in the imaginary part of the propagator, $\Im \Delta_+(p_0,0)$, are shown for $g=2.5$ (upper panel) and $g=4.5$ (lower panel).}
    \label{fig:cuts}
\end{figure}

Starting at small couplings (there is no qualitative change of the figure at couplings smaller than $g=1.5$ as depicted in the figure), we identify three poles on the real axis ($\omega = \Re p_0$), at $\omega= \pm m_q(\gamma_\smallG)$ and at $\omega=0$. We will discuss the features of these poles in more detail in Sec.~\ref{sec:mass}. At larger couplings, e.g. $g=3.5$, the two peripheral poles approach the origin, and at around $g\gtrsim 4.5$, the three poles seem to collapse onto each other. As will be demonstrated more clearly in Sec.~\ref{sec:transition-pfinite}, the GZ pole merges with the plasmino pole, while the plasmon pole persists with vanishing mass. The merging of poles is accompanied by the emergence of additional non-analytic features in the complex plane. At $g=2.5$ the propagator $\Delta_+(p_0,0)$ is continuous apart for developing complex poles at $\Re p_0=0$ and $\Im p_0 =\pm0.1 m_q$ along with branch cuts above and below, also see Fig.~\ref{fig:cuts} (upper panel). In contrast, at $g=4.5$ we only see pronounced discontinuities in the imaginary part of the propagator $\Im \Delta_+(p_0,0)$, see Fig.~\ref{fig:cuts} (lower panel), indicating that the structures seen in the spectral density indeed are branch cuts in the complex plane.

Remarkably, at large couplings the three poles have become one stable pole. Simultaneously, the cuts in the complex plane gradually vanish with the increasing coupling constant. We have checked that the propagator becomes asymptotically analytic with a single pole at $\omega = p=0$ when $g\to \infty$. This clearly indicates the restoration of Lorentz symmetry in the vacuum, which is a highly non-trivial consistency check of the underlying theory as a proper description of strongly coupled QCD that contains an intrinsic non-Abelian transition. As a result of Lorentz symmetry breaking being responsible for the transition, the hydro-like massless mode emerged at small-$g$ uncovered previously~\cite{Su:2014rma} by two of the authors can be identified as a corresponding Goldstone mode following the findings in Ref.~\cite{Ojima:1985is}. This is in close analogy to the phonon as the Goldstone mode of the spontaneously broken Galilean symmetry in fluids~\cite{Leutwyler:1996er}. Furthermore, this massless mode at large $g$ is in surprising agreement with phenomenological predictions from Dyson-Schwinger equations~\cite{Nakkagawa:2011ci,Nakkagawa:2012ip} and gauge/gravity duality~\cite{Seo:2012bz,Seo:2013nva}.

\subsection{From massive to massless excitations}
\label{sec:mass}

Let us now turn in more detail to the study of the poles on the real axis. When $p =0$, the pole located on the real axis $\omega =\Re p_0$ is simply $\omega = \tilde \Sigma(\omega,0)$. In this limit we get 
\begin{equation}
    \tilde \Sigma(\omega,0) = \frac{2g^2 C_F}{(2\pi)^2} {\rm Re} \int_0^\infty \rmd k\, \frac{\tilde n_+(k)\,k^2 \,\omega}{E_+ \left[\omega^2 - (E_+-k)^2\right]} \,.
\end{equation}
This simplified form sheds us some analytic insight on the behavior of the two-point function with the strong coupling constant.

Let us first consider the massless pole. Taking directly the limit $\omega \to 0$ gives us $\tilde \Sigma(\omega,0) = \alpha_\smallG \omega + \mathcal{O}(\omega^3)$, where
\begin{equation}
    \label{eq:alphaG}
    \alpha_\smallG = -\frac{2g^2 C_F}{(2\pi)^2} {\rm Re} \int\rmd k\,\frac{\tilde n_+(k)\, k^2}{E_+(E_+-k)^2} \,.
\end{equation} 
Note that $\tilde \Sigma(\omega,0)$ is purely real. Since we only have two scales available, namely $\gamma_\smallG$ and $T$, and since $\alpha_\smallG$ is dimensionless, it is necessarily only function of $\gamma_\smallG/T$ and, therefore, temperature independent. In this case the propagator becomes
\begin{equation}
    \lim_{\omega \to 0}\Delta_+(\omega,0) = \frac{Z_0}{\omega} \,,
\end{equation}
which, as expected, has a pole at $\omega=0$. The residue of the massless excitation is simply $Z_0 = 1/(1-\alpha_\smallG)$. This result implies there is always a massless pole present in the theory whenever we have a finite $\gamma_\smallG >0$. In contrast, for $\gamma_\smallG=0$ exactly, this pole never occurs because $\Im \tilde \Sigma(\omega,0) \neq 0$ in this regime.

\begin{figure}
    \centering
    \includegraphics[width=\columnwidth]{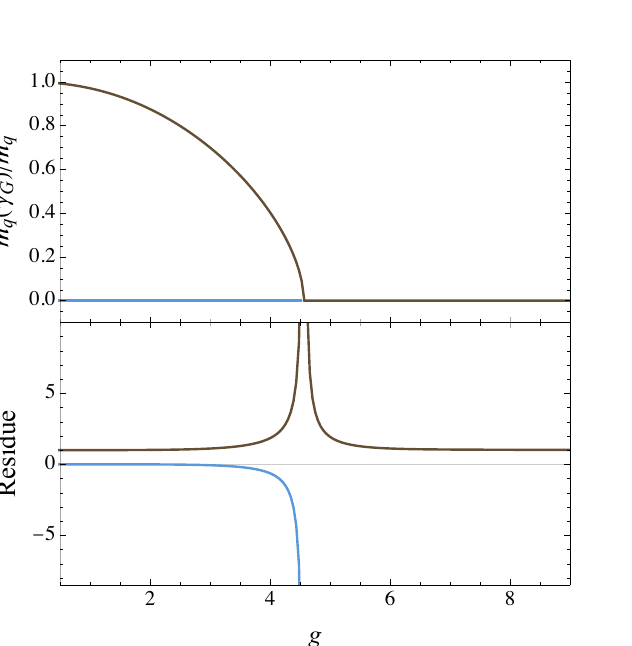}
    \caption{The mass (upper panel) and residues (lower panel) of the quark and anti-quark hole excitations (brown curves) and GZ pole (blue curves) are shown as functions of the coupling $g$. The critical coupling is $g_* \approx 4.562$.}
    \label{fig:zero-mode}
\end{figure}

In addition to the massless pole, there are two more poles at small coupling which belong to the fermion and anti-fermion hole massive excitations of the plasma. We separate out a $\omega^{-1}$ contribution in the propagator by defining $\tilde f(\omega) = 1- \tilde \Sigma(\omega,0)/\omega$. Now the massive pole can be located by $\tilde f(\omega)= 0$. In this limit the propagator reads
\begin{equation}
    \lim_{\omega \to m_q(\gamma_\smallG)} \Delta_+(\omega,0) = \frac{Z_q}{\omega- m_q(\gamma_\smallG) }\,,
\end{equation}
and similarly for the pole at $\omega = -m_q(\gamma_\smallG)$. 
These define the quark (plasmon) $\omega_+ = m_q(\gamma_\smallG)$ and anti-quark hole (plasmino) $
\omega_- = m_q(\gamma_\smallG)$ masses.
Their residues are $Z_q = Z_\pm = 1/[\omega \tilde f'(\omega)]_{\omega= \pm m_q(\gamma_\smallG)}$. In the small coupling limit, i.e. $g \to 0$, we directly obtain $Z_+ = Z_- =1/2$.

To compare our analytic expectations with the numerical evaluation we plot the normalized mass $m_q(\gamma_\smallG)/m_q$ as function of the coupling in the upper panel of Fig.~\ref{fig:zero-mode}. Reflecting the merger of the poles observed in Fig.~\ref{fig:transition-p0}, we see that the thermal quark mass decreases monotonously with the coupling, vanishing at critical coupling $g_\ast$. The massless mode is present in the system at any coupling. The sharpness of the behavior of the thermal mass indicates the presence of a phase transition between a 3-mode and a 1-mode configuration, and $m_q(\gamma_\smallG)$ serves as an order parameter of the transition. The behavior of the massive modes, see brown lines in Fig.~\ref{fig:zero-mode} (also, see the behavior of the speed of sound of the system in Fig.~\ref{fig:vs-abs} that is to be discussed in Sec.~\ref{sec:transition-dofs}), indicates that the phase transition of strongly coupled QCD with massless quarks is 2nd order at apparent level, which surprisingly coincides with the latest lattice observation~\cite{Cuteri:2021ikv}. We will come back to an interpretation of our results in Sec.~\ref{sec:discussion}.

We plot $Z_0$ and $Z_++Z_-$ as function of the coupling in the lower panel of Fig.~\ref{fig:zero-mode}. Strikingly, at small coupling the residue of massless GZ pole is very small $Z_0 \approx 0$ and $Z_+ + Z_- \approx 1$, as expected. Above the transition, $g > g_\ast$, the plasmino mode and the GZ mode merge and cease to exist, hence $Z_0 = Z_- =0$. The only remaining mode is the quark, hence $Z_+=1$ at very large coupling $g$. This further supports the findings in Sec.~\ref{sec:dofs-p0} that the propagator at large coupling contains a simple pole and is otherwise analytic. Around the critical coupling $g_\ast$ we also observe additional complicated features in the complex plane.

We define the critical coupling $g_*$ at the point where $\alpha_\smallG = 1$ or  $Z_0$ diverges, i.e. approaching the critical coupling from below $\lim_{g \to g_\ast} |Z_0| = \infty$. Numerically, we identify the critical coupling to be $g_* \approx 4.562$.

\subsection{Sum rules in the $p\to 0$ limit}
\label{sec:sum-rules-p0}

\begin{figure}
    \centering
    \includegraphics[width=\columnwidth]{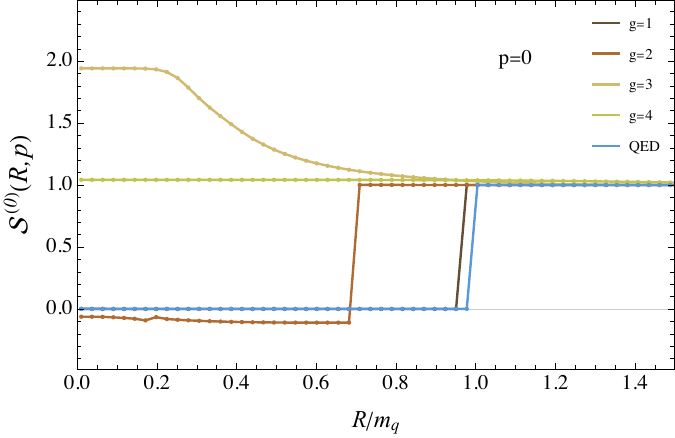}
    \caption{The partial sum rule $\mathcal{S}^{(0)}(R,0)$ for various coupling strengths in QCD, compared with the result in QED.}
    \label{fig:sumrule-p0}
\end{figure}

In this subsection, we demonstrate the sum rule in Eq.~(\ref{eq:sumrule_n=0}) is fulfilled for the analysis as discussed in the previous two subsections. We plot the \emph{partial} sum rule $\mathcal{S}^{(0)}(R,0)$, defined in Eq.~\eqref{eq:partial-sum-rule-n0}, for $n=0$ and with $p=0$ in Fig.~\ref{fig:sumrule-p0}. As expected, both for QED, i.e. $\gamma_\smallG=0$, and for QCD with small coupling constants $g$ we observe a single jump of the function at $R=m_q(\gamma_\smallG)$, corresponding to the two, symmetric poles located exactly at $\omega=\pm m_q(\gamma_\smallG)$. 

Increasing the coupling to $g=3$ we see indications of the onset of the 3-mode $\to$ 1-mode transition and $\mathcal{S}^{(0)}(R,0) >1$ everywhere. The plateau at $0 < R/m_q < 0.3$ is the result of the cut running along (or close to) the imaginary axis observed in Fig.~\ref{fig:transition-p0}. However, it asymptotically tends to unity which implies that the sum rules are always satisfied. 
At increasingly large $g > g_\ast$, the partial sum rule is always identical to 1 everywhere.

\section{QCD transition at finite momenta}
\label{sec:transition-pfinite}

At finite spatial momenta $p>0$, the propagator also contains information about the kinematics of the excitations. In this section, we map out the finite-$p$ behavior of the three excitations, two massive and one massless, that we discussed in the previous section. We again utilize the sum rule in Eq.~(\ref{eq:sumrule_n=0}) as the non-trivial criterion to guide our study in the same spirit as the last section, and demonstrate its fulfillment in Sec.~\ref{sec:sum-rules-pfinite}.

\subsection{Transmutation of degrees of freedom at finite-$p$}
\label{sec:transition-dofs}

\begin{figure}
    \centering
    \includegraphics[width=\columnwidth]{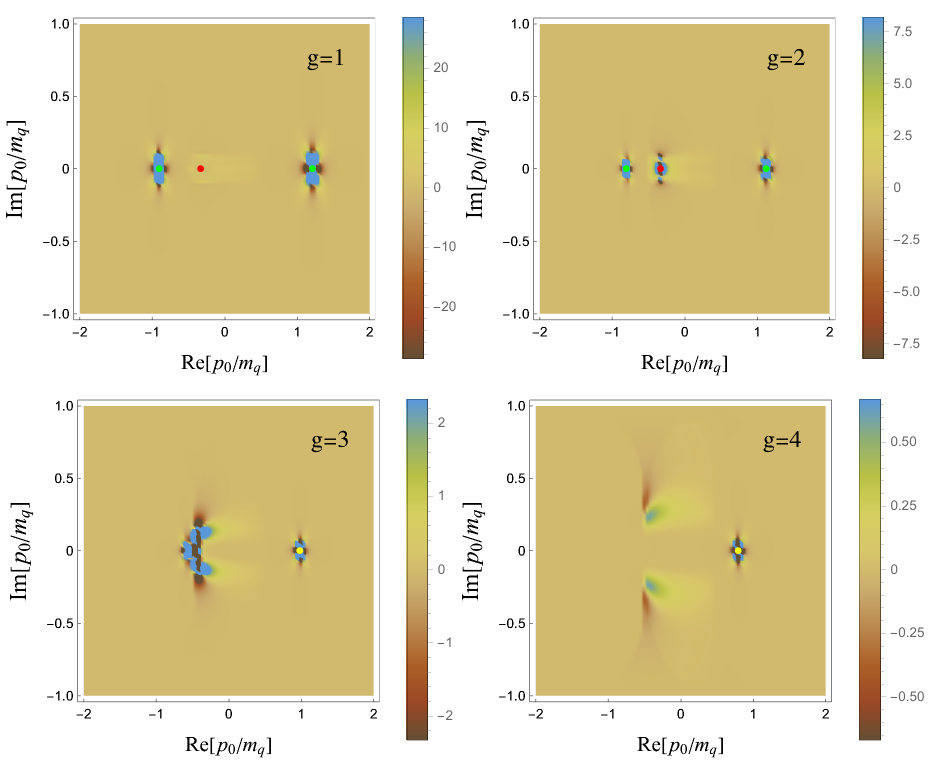}
    \caption{Plots of the real part of the spectral density at $p=0.5 m_q$, $\Re d_+(p_0,p)$, for a range of coupling strengths. The locations of poles on the real axis have been marked with red (the GZ pole) and green (the plasmon and plasmino poles) dots at small couplings, while the unique pole is marked with white dot at large couplings.}
    \label{fig:transition-finitep}
\end{figure}

We plot the spectral density with $p = 0.5\, m_q$ for a wide range of couplings in Fig.~\ref{fig:transition-finitep}. Remarkably, the same features are qualitatively repeated as in Fig.~\ref{fig:transition-p0}: i) non-analyticities in the complex plane and ii) a $3 \to 1 $ transition in terms of collective degrees of freedom.

At small coupling we observe three poles located on the real axis. As before, the pole at positive energy $\omega = \omega_+(p)$ is the dressed, thermal quark (green point). At $\omega<0$ we now find two poles: the leftmost pole $\omega = \omega_-(p)$ is the familiar anti-quark excitation (plasmino) in the medium (green point). Finally, we also have a third pole in the space-like regime $|\omega| < p$ which we denote $\omega = \omega_\mathrm{\smallGZ}$ (red point). This is the remnant of the massless pole observed at $p=0$ in Sec.~\ref{sec:transition-p0}. We discuss the dispersion relations of each of these poles in Sec.~\ref{sec:dispersion-relations}. 

\begin{figure}[t]
\centering
\includegraphics[width=\columnwidth]{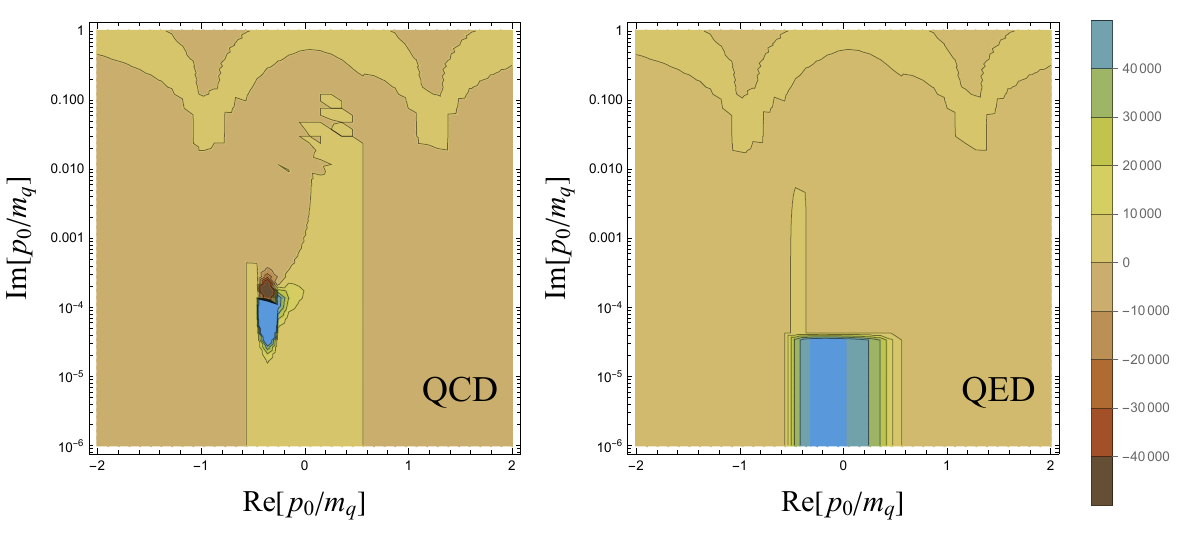}
\caption{An enhanced view of the real part of the spectral density, $\Re d_+(p_0,p)$, for QCD (left panel) and QED (right panel) close to the real axis for $p=0.5 m_q$ and $g=0.3$.  In QED, the spectral density exhibits a branch cut extending over $-p < \Re[p_0] < p$, which is suppressed in QCD, leaving a contribution to the non-analytic structure at finite $\Im[p_0]$.}
\label{fig:spectral-zoom}
\end{figure}

Note that the cut along the real axis spanning $-p < \omega < p$, expected from the QED propagator, has vanished (Fig.~\ref{fig:transition-finitep}, upper-left panel). In order to shed light on this intriguing feature, we zoom in on a region close to the real axis, for $\Im p_0>0$, at $g=0.3$ in Fig.~\ref{fig:spectral-zoom} where the imaginary axis is plotted on a log-scale. As a result, the poles on the real-energy axis are not visible in the plot. For QED, Fig.~\ref{fig:spectral-zoom} (right panel), we see clearly an emerging region of non-analytic behavior in the spectral density spanning $-p < \omega < p$ with $p = 0.5\, m_q$. This is the familiar cut leading to Landau damping. In the QCD case, Fig.~\ref{fig:spectral-zoom} (left panel), there is no such extended region. Instead, we see a non-analytic feature localized close to, but distinctively away from, the real axis. This provides further evidence that there is no analog of Landau damping in this situation. We note that the non-analytic feature is located very close to the massless pole.

Returning to Fig.~\ref{fig:transition-finitep}, when the coupling increases, e.g., $g=3\sim4$, the anti-plasmino and the GZ poles merge and cancel out on the real axis, e.g., in Fig.~\ref{fig:transition-finitep} (lower, left panel). This is accompanied by pronounced and extended non-analytic regions in the complex plane which gradually leave the real axis. There is however no discontinuity of $\Delta_+(p_0,p)$ involved; these are regions of smooth but non-analytic behavior. At larger coupling $g>g_*$, all non-analytic features are gradually suppressed in the complex plane, except of a single pole, located firmly on the real axis at $\omega = p$, which again confirms the restoration of Lorentz symmetry.

Note that the merger of the poles occurs at a coupling distinctly different from $g_\ast$, which was identified at $p=0$. We will return to this point below in Sec.~\ref{sec:dispersion-relations}. These features are strikingly different from QED and are fundamentally encoded in the structure of the resummed quark propagator.

\subsection{Dispersion relations}
\label{sec:dispersion-relations}

To have a better understanding of the intrinsic transition from three modes to one mode with the increasing coupling, we analytically explore the dispersion relation of these collective excitations in detail. Since we are interested in poles on the real axis, we only consider $\omega = \Re p_0$, and expand the self-energy around small $p$ to obtain
\begin{multline}
    \label{eq:sigma-tilde-full}
    \tilde\Sigma(\omega,p) = \frac{g^2 C_F}{(2\pi)^2} \Re \int \rmd k\, \tilde n_+(k) \Big\{ \frac{2 k^2 \omega}{E_+ \left[\omega^2 - (E_+-k)^2\right]} \\ - \frac{2k^3}{3 E_+^2}\frac{\omega^2 +(E_+-k)^2}{\left[\omega^2- (E_+-k)^2\right]^2} p \Big\} + \mathcal{O}(p^2) \,.
\end{multline}
Taking directly $\gamma_\smallG \to 0$ at this point, leads to the QED soft propagator, 
\begin{equation}
    \Delta_+^{-1}(\omega,p) = \omega \left(1-\frac{m_q^2}{\omega^2} \right) - p \left(1 - \frac{m_q^2}{3 \omega^2} \right) \,.
\end{equation}
This has poles at $\omega_\pm(p) = m_q \pm \frac13 p$, corresponding to the standard fermionic modes, the quark and the plasmino.

\begin{figure}
    \centering
    \includegraphics[width=\columnwidth]{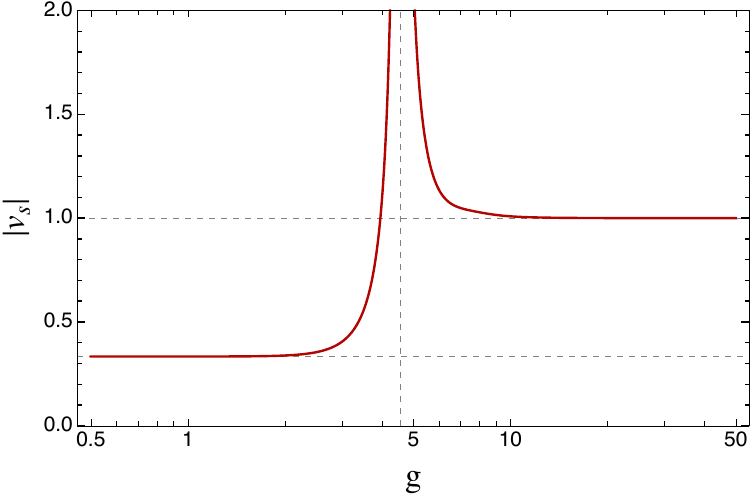}
    \caption{The absolute value of the speed of sound, $v_s$, for soft/static modes is shown as a function of the coupling $g$. At small coupling, $v_s$ is negative, approaching $-1/3$, while at large coupling, $v_s$ becomes positive and asymptotically approaches $1$. At critical coupling, $g_* \approx 4.562$, $v_s$ is undefined, indicated by a dashed vertical line.}
    \label{fig:vs-abs}
\end{figure}

The self-energy in Eq.~\eqref{eq:sigma-tilde-full} describes a novel massless excitation when taking the $\omega \to 0$ limit. In this case, we need to expand $\tilde \Sigma(\omega,p)$ to order $\mathcal{O}(\omega)$ and $\mathcal{O}(p)$. Avoiding technical details, we get that $\tilde \Sigma(\omega,p) = \alpha_\smallG \omega + \beta_\smallG p$, where $\alpha_\smallG$ was already given in Eq.~\eqref{eq:alphaG} and 
\begin{equation}
    \beta_\smallG = -\frac{2g^2 C_F}{3(2\pi)^2} \Re \int_0^\infty \rmd k\,  \frac{k^3\, \tilde n_+(k)}{E^2_+(E_+-k)^2}\,.
\end{equation}
This results in the propagator
\begin{equation}
    \label{eq:inv-prop-soft}
    \Delta_+(\omega,p) = \frac{Z_0}{\omega - v_s p} \,,
\end{equation}
where $Z_0 = 1/(1-\alpha_\smallG)$ is the residue of the massless pole at $p\to0$ limit and the speed of sound $v_s$ of the system is
\begin{equation}
    \label{eq:vs}
    v_s = \frac{1+\beta_\smallG}{1-\alpha_\smallG} \,.
\end{equation}
The dispersion relation of the massless pole is therefore $\omega(p) = v_s p$, where $v_s$ depends only on the coupling constant.

\begin{figure}
    \centering
    \includegraphics[width=\columnwidth,trim={0 0.25cm 0.5cm 1.1cm},clip]{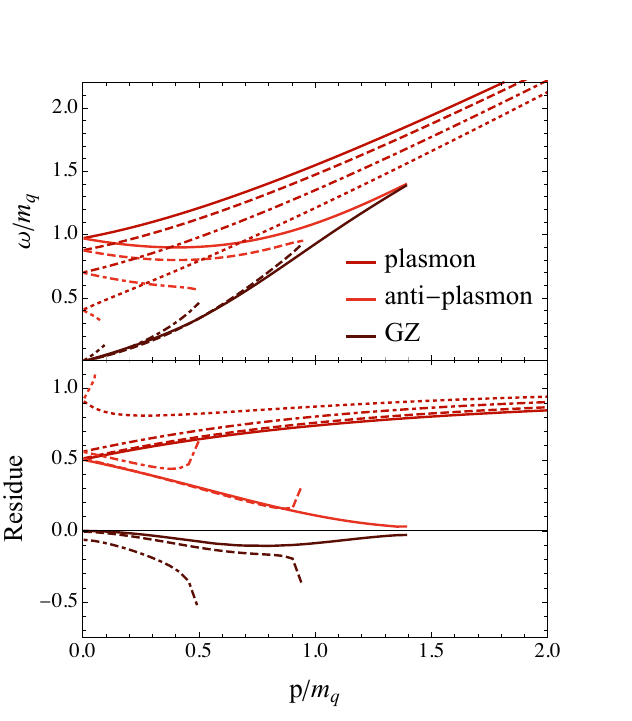}
    \caption{Dispersion relations (upper panel) and residues (lower panel) for poles on the real axis are shown for $g=1$ (solid curves), $g=2$ (dashed curves), $g=3$ (dot-dashed curves) and $g=4$ (dotted curves).}
    \label{fig:dispersion-residues}
\end{figure}

At small coupling the pole of Eq.~\eqref{eq:inv-prop-soft} is at negative $\omega$, and the speed of sound of the system is $|v_s|=1/3$ at $g < 2$. This pole cannot therefore be associated with a quasi-particle excitation in the medium because it is space-like, i.e. $\omega < p$, see Fig.~\ref{fig:vs-abs} for the speed of the sound of the soft modes in the two phases. Adding to its unconventional nature, we notice that its residue $Z_0(p)$ is negative, see Fig.~\ref{fig:dispersion-residues} (lower panel, brown lines) \cite{Su:2014rma}. At $g = g_\ast$, the speed of sound is not well defined, but in the large coupling regime the pole moves to positive $\omega$ and the speed of sound becomes unity, i.e. $\omega=p$, see Fig.~\ref{fig:vs-abs}.

The full dispersion relations $\omega_\pm(p)$ and $\omega_\mathrm{\smallGZ}(p)$, reconstructed numerically, are plotted in Fig.~\ref{fig:dispersion-residues} (upper panel) for the four chosen couplings $g=1$, $g=2$, $g=3$ and $g=4$, together with their respective residues (lower panel). As seen in Fig.~\ref{fig:transition-finitep}, we see that the plasmino and GZ poles merge at a finite $p$, which decreases with increasing coupling strength $g$. The remaining mode becomes more and more light-like.

The dispersion relations in the weak coupling limit were first discussed in \cite{Su:2014rma}, where it was noted that the non-analytic features on the real axis, i.e. the residues, do not add up to 1, i.e. $Z_0(p) + Z_q(p) + Z_{\bar q}(p) < 1$. The additional contributions needed to satisfy the sum rule are found in the complex plane, see Fig.~\ref{fig:spectral-zoom}.

\subsection{Sum rules at finite momenta}
\label{sec:sum-rules-pfinite}

\begin{figure}
    \centering
    \includegraphics[width=\columnwidth]{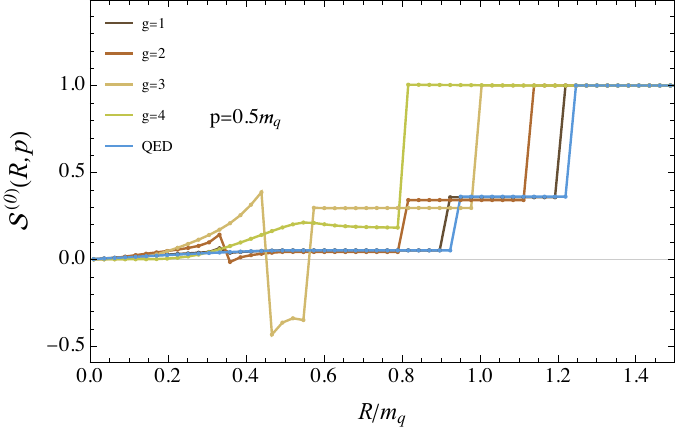}
    \caption{The partial sum rule $\mathcal{S}^{(0)}(R,p)$ at $p=0.5 m_q$ for  various coupling strengths in QCD, compared with the result in QED.}
    \label{fig:sumrule-finitep}
\end{figure}

In Fig.~\ref{fig:sumrule-finitep}, we plot the partial sum rule in Eq.~(\ref{eq:sumrule_n=0}) as a function of the radius $R$ in the complex plane for several coupling strengths corresponding to Fig.~\ref{fig:transition-finitep} and for QED (blue curve). Focusing on the latter, there is a small contribution to the sum rule from the space-like cut, followed by two larger jumps accounting for the massive poles. At large $p_0$, the function is fully analytic and the sum rule is satisfied.

The situation is dramatically different for QCD at large coupling. In addition to the migration of the poles and their subsequent mergers, we observe that the smooth but non-analytic features in the complex plane are quite intricate and can cause $\mathcal{S}^{(0)}(R,p)$ to vary rapidly and even become negative. Reassuringly, at a sufficiently large radius the propagator is again fully analytic and the sum rule is restored for any coupling $g$. This is a crucial consistency check for the whole formalism and a necessary prerequisite that allows us to study the theory from weak to strong coupling.

\section{Discussions and conclusions}
\label{sec:discussion}

In this paper, we have established for the first time spectral sum rules for strongly coupled QCD at finite temperature, Eqs.~\eqref{eq:sumrule_n=0}-\eqref{eq:sumrule_n=2}, by incorporating the non-perturbative magnetic scale via the GZ quantization. The first, and most basic, one is required by quantum mechanics, see Eq.~(\ref{eq:sumrule-qm}). The associated quark propagator features smooth but non-analytic regions in the complex plane that have, to our knowledge, not been observed before. These newly discovered regions confirm that strongly coupled quarks obey quantum statistics as fermions, establishing a key consistency condition for GZ quantization to serve as a systematic framework for describing strongly coupled QCD at finite temperatures. In stark contrast to the fermionic spectral functions in QED and weakly coupled QCD  that are one-dimensional in the complex plane, the strongly coupled quark spectral function features two-dimensional structures in the complex-energy plane, i.e $p_0 \in \mathbb{C}$. This insight could have significant implications for extracting physical information from lattice QCD data. In order to map out these features, we have introduced a novel quantity, the spectral density, that is specifically designed to pick up non-analytic features in the complex-energy plane. 

A QCD transition has been identified in light of the sum rule in Eq.~\eqref{eq:sumrule_n=0}, revealing remarkable intrinsic non-Abelian features, see Fig.~\ref{fig:transition-p0} for vanishing and Fig.~\ref{fig:transition-finitep} for finite spatial momentum, respectively. As the coupling $g$ increases, the system undergoes a transition from a phase with two massive time-like modes and one massless space-like mode with positive violation to a phase with only one massless light-like mode in the asymptotic limit. This transition is induced by the magnetic scale, encoded in $\gamma_\smallG$. 

Furthermore, Lorentz symmetry breaking is responsible for the transition with the space-like massless mode at small-$g$ being the corresponding Goldstone mode. The thermal mass vanishes across this weak-to-strong $g$ transition, see Fig.~\ref{fig:zero-mode}, which is in close agreement with various phenomenological predictions of strongly coupled QCD~\cite{Nakkagawa:2011ci,Nakkagawa:2012ip,Seo:2012bz,Seo:2013nva} and serves as a direct order parameter for the phase transition, whose order coincides with the latest lattice result~\cite{Cuteri:2021ikv}. It is very surprising to see one formalism to accommodate all these non-perturbative QCD effects in a systematic manner, provided $g$ is the only parameter as required by the QCD theory.

All the results in our setup are genuine non-Abelian phenomena induced by the non-perturbative magnetic scale. They demonstrate a rich and sophisticated phase structure of strongly coupled QCD originating from the quantum statistics of fundamental particles, which has shed new light on the emergence and mechanism of the QCD deconfinement transition. The application of GZ quantization has shown novel phenomenological features of a strongly coupled QCD plasma~\cite{Kojo:2009ha,Kojo:2011cn,Kojo:2012js,Bandyopadhyay:2015wua,Florkowski:2015rua,Hattori:2015aki,Florkowski:2015dmm,Begun:2016lgx,Jaiswal:2020qmj,Madni:2022bea,Sumit:2023hjj,Debnath:2023dhs,Bandyopadhyay:2023yjp,Debnath:2023zet,Sumit:2023oib,Madni:2024xyj,Madni:2024ubw}. Furthermore, the method developed here may help to understand similar strongly coupled systems, such as the quantum spin liquid~\cite{stone2006quasiparticle,lee2007high,broholm2020quantum}.

\section*{Acknowledgments}
This paper is dedicated to the memory of the late Daniel Zwanziger (1935--2024).  We acknowledge Agnes Mocsy for discussions during QM2006 that stimulated the early idea of this work. We thank M.~Chernodub, W.J.~Fu, J.M.~Pawlowski, H.~St\"ocker, M.~Strickland and A.~Rothkopf for inspiring discussions. This work is supported by the Taishan Scholars Program (No. tsqnz20221162) and Shandong Excellent Young Scientists Fund Program (Overseas) (No. 2023HWYQ-106).


\bibliography{main_v1}

\end{document}